# Hypersonic properties of monodisperse spherical mesoporous silica particles

D A Eurov[1], D A Kurdyukov[1], E Yu Stovpiaga[1], A S Salasyuk[1], J Jäger[2], A V Scherbakov[1], A V Akimov[1,3], A J Kent[3], D R Yakovlev[1,2], M Bayer[1,2] and V G Golubev[1]

[1]Ioffe Physical-Technical Institute, 194021 St. Petersburg, Russia
[2]Experimentelle Physik 2, Technische Universität Dortmund, D-44227 Dortmund, Germany
[3]School of Physics and Astronomy, University of Nottingham, NG7 2RD Nottingham, United Kingdom

Email: edan@mail.ru

**Abstract.** We use the picosecond acoustic pump-probe technique to study the elastic properties of monodisperse mesoporous silica spheres filled with nickel and deposited in the form of opal-like films on silica substrates. The picosecond pump-probe optical transmission signal shows harmonic oscillations corresponding to the lower energy radial Lamb mode in the vibrational spectrum of the spheres. These oscillations, with a frequency of several gigahertz last for several nanoseconds in the spheres with diameter 1050 nm, showing high homogeneity of the sphere parameters. By analysis of the oscillation spectrum of films with different sphere diameter and nickel content we obtain the elastic moduli of the mesoporous silica spheres.

The synthesis and study of mesoporous silica materials with unique and advantageous properties, such as highly ordered porous structure, large surface area and pore volume, narrow pore size distribution, and tunable size and structure of pores have been attracting intensive attention. Prominent research interest has been triggered by the wide range of their prospective applications in adsorption, separation, catalysis, sensing, energy conversion and storage, biotechnology and nanomedicine fields. Significant advances have been made in their structural, compositional, and morphological control and functionalization [1-11].

Monodisperse spherical mesoporous silica particles (MSMSPs), with a size standard deviation not exceeding 10%, and highly ordered mesoporous structures with controllable mean pore and particle sizes would further extend the spectrum of applications. Among prospective applications are: fabrication of magnetic nanocomposites [12]; capillary electrochromatographic enantioseparation [13]; using MSMSPs as high-load-capacity containers for anticorrosive organic inhibiting agents [14]; color control of photonic-crystal films [15, 16]; and slowing down the spontaneous emission in laser systems [17,18]. The possibility of controlling the size of nanoparticles in the range of 50-1000 nm and morphology of MSMSPs is very appealing for biomedical science and technology. MSMSPs have been used as drug-delivery vehicles, cell imaging agents, cell markers, carriers of molecules, and means of bioanalysis [19-22]. In novel targeted drug delivery systems, MSMSPs could improve control over the spatial and temporal kinetics of drug release at the site of action to achieve the optimal pharmokinetic effect.

Among the numerous methods and applications with MSMSPs and MSMSP-based nanocomposites, ultrasonic techniques play an important role. Being invaded into biological tissue, porous silica particles can be used as a contrast agent for ultrasonic imaging in medicine [23]. Monodispersity and tailorable size will allow improved targeted administration of the particles at the tumor site. However there is very little information about the nanomechanical and elastic properties of MSMSPs. Studies of these properties in various nanostructured materials have been always important for various applications [24], and this information would be extremely valuable for ultrasonic and photoelastic imaging of MSMSPs embedded into biological tissues. At first sight the elastic properties of MSMSPs can be considered in a similar way as for solid, not porous, silica spherical particles [25,26], but with reduced longitudinal (LA) and transverse (TA) sound velocities $s_L$ and $s_T$, respectively [27-30]. However the elastic modulus and Poisson's ratio for the MSMSPs depend on the morphology of the pores and are usually unknown unless studied in the exactly the same bulk porous material. Another experimental challenge is related to the high frequency of localized vibrational modes of MSMSPs. For micrometer and sub-micrometer values of the sphere diameter $D$ the values of the vibrational frequencies fall into the hypersonic GHz and sub-THz frequency ranges respectively which cannot be easily reached by standard ultrasonic methods so that more sophisticated hypersonic pump-probe techniques should be used [31-36]. The application of laser ultrasonic pump-probe techniques to bulk porous materials has been demonstrated recently [37,38] but no hypersonic studies of MSMSPs has been done.

The main task of the present work is to study the hypersonic vibrations in MSMSPs by picosecond acoustic techniques and measure the frequency and decay of localized vibrational modes excited as a result of absorption of the pump light by the metal inclusions in the pores of MSMSPs. We use an ensemble of MSMSPs with very low dispersion of $D$ which allows us to measure the quality ($Q$-factor) of the localized mode and obtain values for the elastic modulus in MSMSPs.

MSMSPs were synthesized by a soft template method using cetyltrimethylammonium bromide (CTAB) as structure-directing agent. The particles form in the course of tetraethyl orthosilicate (TEOS) hydrolysis in a CTAB-ethanol-water-ammonia mixture [39,40]. To remove organic substances, the synthesized particles were annealed in air at 550 °C. The value of the true specific gravity of the particle's material – amorphous $SiO_2$, measured with a helium pycnometer, was 2.1 g cm$^{-3}$. MSMSPs with mean diameters of 620 and 1050 nm were synthesized. The standard deviation of particle diameters was less than 4%. The diameter of the mesopores and specific surface area were 3.1±0.15 nm and around 800 m$^2$ g$^{-1}$ for MSMSPs with both $D$. The pore volume of the particles with 620 and 1050 nm in diameter was 0.56 cm$^3$ g$^{-1}$ (54% of particle's volume) and 0.63 cm$^3$ g$^{-1}$ (57 vol.%), respectively.

To study the hypersonic properties of the MSMSPs, they were prepared in the form of closed-packed opal-like films. Such an approach was recently applied for monodisperse spherical solid silica particles [41,42]. Using films of highly ordered MSMSPs instead of the disordered ensemble of nanoparticles is essential for using the picosecond acoustics pump-probe technique which requires good optical transmission and reduced Rayleigh scattering. Films consisting of seven MSMSPs monolayers were fabricated from 1 wt.% MSMSPs water suspension by a vertical deposition technique [43] on substrates made from optically polished microscope glass of the Teget brand. After film fabrication the MSMSPs were filled with Ni: at first, the pores of the colloidal films were filled with 1M solution of $Ni(NO_3)_2 \cdot 6H_2O$ by the lateral infiltration technique [44]; after that the films were annealed at 200 °C, as a result of which nickel nitrate decomposed into oxide. This procedure was repeated cyclically in order to achieve the needed pore filling degree. At the next step, the samples filled with nickel oxide, were annealed at 500 °C under hydrogen flow and as a result the MSMSPs were filled with Ni. The pore filling degree with Ni was evaluated from the particle's density value calculated by the Stokes equation. The sedimentation velocity of particles that is needed for the calculation was measured experimentally. The parameters of the five MSMSP samples used in the present work are presented in table 1. The samples are labeled according to the value of $D$ (1 or 2) and Ni filling factor (A, B, or C) and their parameters are shown in table 1.

The experiments were carried out at room temperature. The experimental scheme, shown in figure 1, is based on the femtosecond pump-probe optical technique. The beam from a femtosecond laser with regenerative amplifier (wavelength 800 nm, pulse duration 200 fs, pulse maximum energy 2 µJ, repetition rate 100 kHz) is split into pump and, second harmonic (400 nm), probe beams. The beams were focused at the surface of the sample, overlapping their spots. The diameters of the focused pump and probe beams were 150 and 50 µm respectively. The pump was incident at 40° angle relative to the surface of the sample and the probe had normal incidence. The maximum energy density of the excitation did not exceed 10 mJ cm$^{-2}$. The experimental signals were measured as relative intensity changes of the transmitted probe beam $T(t)$ as function of time delay between pump and probe, provided by a variable delay line. In order to improve the signal to noise ratio the pump beam was modulated by a 2 kHz mechanical chopper and the transmitted probe beam was detected by a photodiode, the output of which was connected to a lock-in amplifier referenced to the chopper.

**Table 1.** Structural and elastic parameters of the studied MSMSPs (the Poisson's ratio is assumed to be constant and has the same value as in bulk silica, $v=0.17$). Values in brackets are calculated from Eq. (3).

| Sample | Sphere diameter, $D$ (nm) | Porosity, $p$ (% vol.) | Ni filling factor (% from pore vol.) | Density of single sphere, $\rho$ (g cm$^{-3}$) | Frequency of the Lamb mode, $f_0$ (GHz) | Sound velocities $s_L$ (km s$^{-1}$) | $s_T$ (km s$^{-1}$) | Bulk modulus, $B$ (GPa) | Shear modulus, $G$ (GPa) | $Q=f_0\tau$ |
|---|---|---|---|---|---|---|---|---|---|---|
| 1A | 1050 | 57 | 3 | 1.05 | 3.53 | 4.85 | 3.06 | 11.6 | 9.83 | 4.8 |
| 1B | 1050 | 57 | 7 | 1.25 | 3.35 | 4.60 | 2.90 | 12.4 | 10.5 | 4.7 |
| 1C | 1050 | 57 | 25 | 2.15 | 3.28 | 4.51 | 2.85 | 20.7 | 17.5 | 1.5 |
| **Bare porous sphere** | **1050** | **57** | **0** | **0.90** | **3.68**[a] | **5.06**[b] | **3.19**[b] | **10.8 (10.6)** | **9.16** | **~5** |
| 2A | 620 | 54 | 3 | 1.10 | 6.24 | 5.06 | 3.19 | 13.2 | 11.2 | 1.6 |
| 2B | 620 | 54 | 9 | 1.41 | 6.01 | 4.88 | 3.08 | 15.8 | 13.4 | 1.7 |
| **Bare porous sphere** | **620** | **54** | **0** | **0.95** | **6.35**[a] | **5.15**[b] | **3.25**[b] | **11.8 (11.5)** | **10.0** | **<2** |
| Solid silica | 1050 620 | - | - | 2.20 | 4.89[a] 8.28[a] | 5.97 | 3.75 | 36.9 | 31.0 | n/a |

[a] Calculated using the values for $s_L$
[b] Extrapolated from the dependence of $s_{L,T}$ on the Ni filling factor

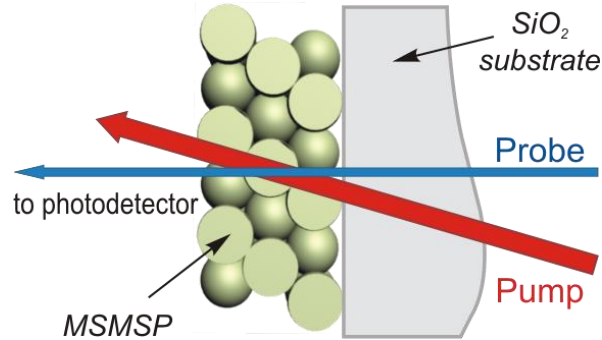

**Figure 1.** Experimental scheme.

Figures 2 (a) and 3 (a) show the temporal evolution of $T(t)$ measured for the samples with $D=1050$ nm and 620 nm respectively. It is clearly seen that the signals show oscillatory behavior. The different curves in figures 2 (a) and 3 (a) correspond to the spheres with various fillings by Ni. The highest amplitude of $T(t)$ is observed in sample 1C which has maximum filling fraction of Ni in the samples with the spheres $D=1050$ nm. The fast Fourier transforms (FFT) of $T(t)$ are presented in figures 2 (b) and 3 (b) from which it is seen that the spectra have one spectral line with a maximum around 3.4 GHz for spheres with $D=1050$ nm and 6.0 GHz in the sample with $D=620$ nm. The frequency corresponding to the maximum of the spectrum slightly decreases with increasing Ni fraction (from 3.5 GHz to 3.3 GHz in the samples with $D=1050$ nm). The width of the spectrum for the spheres with $D=1050$ nm has a value of 0.6 GHz for low filling but doubles in the sample with the highest filling value. Moreover, the spectrum for sample 1C becomes strongly asymmetric showing a wing which spreads towards low frequencies. The samples 2A and 2B with $D=620$ nm also show a single peak in the FFTs (see figures 2 (b) and 3 (b)), and it is centered at a frequency around 6 GHz.

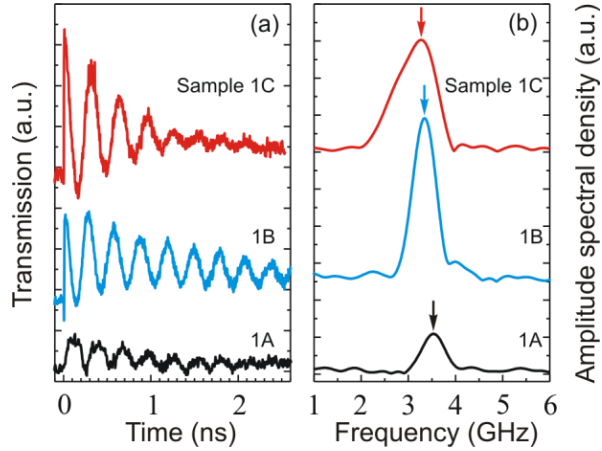

**Figure 2.** (a) Transmission of the probe light as a function of time delay between pump and probe pulses measured in the samples with 1050 nm sphere diameter. The different curves correspond to different nickel content in the silica sphere pores (see Table 1). Panel (b) shows fast Fourier transforms obtained from the measured signals in (a). The curves in (b) are splined. Vertical arrows indicate the frequencies of the maxima in the spectra.

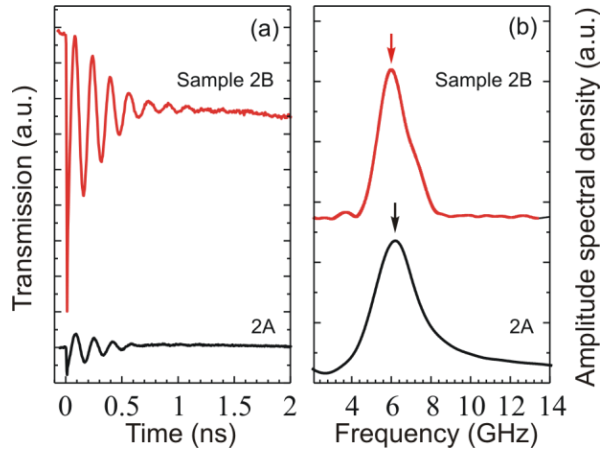

**Figure 3.** (a) Transmission of the probe light as a function of time delay between pump and probe pulses measured in the samples with 620 nm sphere diameter. The different curves correspond to different nickel content in the silica sphere pores (see Table 1). Panel (b) shows fast Fourier transforms obtained from the measured signals in (a). The curves in (b) are splined. Vertical arrows indicate the frequencies of the maxima in the spectra.

The measured oscillations in $T(t)$ correspond to coherent vibrations of the MSMSPs. The optical pump pulse is absorbed by the Ni inclusions in the pores which leads to an increase of the temperature in the spheres and almost instantaneous generation of stress. The TEM images of the studied MSMSPs show that Ni is homogeneously distributed in the spheres [45] and dynamical stress excitation in the spheres should take place with spherical symmetry. Then only the fully symmetric, radial vibrational mode is excited in each sphere. This statement is supported by the existence of only one spectral line in the FFTs of the measured signals in the samples with low fraction of Ni in the pores.

The measured values of the vibrational frequency $f_0$, presented in table 1, are used to determine the elastic parameters of the MSMSPs. In general, the spherical elastic modes in isolated spheres are characterized by two integers $l \geq 0$ and $n \geq 1$ and for fixed Poisson's ratio, $v$, the mode frequency $f_{l,n}$ is proportional to the inverse sphere diameter $D$. The equations for deriving $f_{n,l}$ were first published by *Lamb* in 1882 [46] and the frequency $f_0$ for the lowest energy symmetric radial mode ($l=0$, $n=1$) may be obtained from the transcendental equation which includes the zero and first order spherical Bessel functions $j_0$ and $j_1$:

$$k_T^2 j_0(k_L) - 4k_L j_1(k_L) = 0, \qquad (1)$$

where $k_L = \pi f_0 D / s_L$ and $k_T = \pi f_0 D / s_T$, ($s_L$ and $s_T$ –longitudinal and transverse sound velocities in MSMSP respectively). Eq. (1) allows us to obtain the values of $s_L$ and $s_T$ for known Poisson's ratio and measured value of $f_0$. The obtained values of sound velocities can be used for deriving the values for bulk and shear moduli, $B$ and $G$:

$$s_L = \sqrt{\frac{B + \frac{4}{3}G}{\rho}} \text{ and } s_T = \sqrt{\frac{G}{\rho}}, \qquad (2)$$

where $\rho$ is the density of a single MSMSP. Further we assume that the Poisson's ratio, $v$, is independent on porosity and equal to the value for bulk fused silica, $v=0.17$. This assumption is reasonable for isotropic porous solids [28]. Then, independently on porosity, $p$, the solution of (1) gives $k_L=2.4$ and $k_T=3.8$. The results for $s_L$, $s_T$, $B$ and $G$ are presented in table 1 for the five studied samples. It is seen that the sound velocities $s_L$, and $s_T$, in MSMSPs are lower than in solid silica, but their dependence on the Ni content is very weak.

Most important for applications is to know the parameters in MSMSPs without Ni. These values are obtained using linear extrapolation of the measured $s_L$, and $s_T$ in MSMSPs with different Ni content. Then the values of $B$ and $G$ are obtained using (2). Such extrapolation is reasonable due to the very small dependence of the sound velocities on the Ni content in the studied MSMSPs. We compare the experimentally obtained value for the bulk modulus $B_p$ in MSMSPs with the theoretical calculations using the equation (1) in ref. [38] obtained using the Hashin–Shtrikman approach [47]:

$$B_p = \frac{(1-p)B_0}{1 + p(3B_0/4G_0)}, \qquad (3)$$

where $B_0$ and $G_0$ are the bulk and shear modules in solid silica (see table 1). The values calculated using Eq. (3) and shown in brackets in Table 1 are in excellent agreement with the values of $B_p$ obtained experimentally. The precise comparison of $G_p$ with theory is not relevant in our work because the experimental values were obtained using Eq. (2) together with the assumption of a constant Poisson's ratio $v$ which is not likely to be valid exactly for theoretical model considerations when the pores have specific forms.

Now we turn to the discussion of the decay of the oscillations in the measured $T(t)$. To obtain the decay time, $\tau$, the slow background was subtracted from the measured $T(t)$ and the temporal curves were fitted by the function $\sim \exp(-t/\tau)\cos(2\pi f_0 t)$. The values of the quality factor, defined as $Q = f_0 \tau$ for the studied MSMSPs are presented in the last column of table 1. It is seen that the maximum value of $Q \approx 5$ is measured in the MSMSPs with $D=1050$ nm and low Ni content. We may reasonably extend this result to the bare MSMSPs with large $D$ not filled with Ni and conclude that their $Q$ will not be lower. Such a large value of $Q$ was not measured in earlier works containing ensembles of solid silica spheres [41,42,48].

There are two main mechanisms, which govern the decay of the coherent oscillations in the ensemble of MSMSPs. The first is related to inhomogeneities in the MSMSPs and particularly the dispersion in $D$. The second is due to the loss of coherence on the strongly inhomogeneous elastic bonds between the spherical particles [42]. It is difficult to say what mechanism limits the maximum measured value $Q≈5$, but the decrease of $Q$ with the increase of Ni content can be explained by the increase of bond stiffness between the particles due to some amount of Ni deposited during the infiltration procedure in the regions where the spherical particles adjoin. The stiffening of the elastic bonds between the spheres for high Ni content results in an increasing role of the vibrational modes of sphere complexes which have lower frequency relative to single spheres. This is the most likely reason for appearance of the low frequency wing in the FFT spectrum [see the spectrum for sample 1C in figure 2 (b)]. This fact is in agreement with the result for opal films where the sintering between the spheres is essential [41,42]. The value of $Q≈2$ is smaller for samples with $D=620$ nm and $Q$ does not depend on the Ni content. Apparently the decay of oscillations in these MSMSPs is governed by the dispersion of $D$.

In conclusion we have measured the elastic parameters of mesoporous silica nanoparticles. The experimentally obtained value for the bulk modulus is in good agreement with the value calculated theoretically. The picosecond acoustic experiments show resonant coherent vibrations with frequencies of several GHz and quality factors ~5 for micrometer-sized silica spherical particles. The high quality factor indicates the monodispersed character of the studied particles.

Among the possible applications of monodispersed mesoporous nanoparticles, we would like to underline the infiltration of MSMSPs into biological tissues and cells. Hypersonic studies of cells are already successfully carried out and promise an advantage in submicrometer imaging and corresponding medical diagnostics [31]. Infiltration of MSMSPs into cells for targeted drug delivery and diagnostics is also developed nowadays. Thus, the performance of hypersonic studies of the biological tissues and cells with embedded MSMSPs would allow measurement of the frequency of the resonant vibrations and obtain information about the elastic properties of MSMSPs with the materials infiltrated into the pores. These materials may be drugs for the cell treatment/diagnostics or various kinds of absorbers which are necessary to get optical, NMR, microwave, X-ray or acoustic images. Measurements of the resonant vibration decay can provide information about the elastic contrast with the surrounding media (e.g. cell tissue). In this respect the small dispersion of the MSMSPs becomes extremely important for having long living vibrations in the isolated MSMSPs.

**Acknowledgements**

The authors are grateful to V.V. Sokolov for measurements of the true density of the samples. A. S. Salasyuk and A. V. Scherbakov acknowledge the support of the Russian Academy of Sciences.

**References**

 [1] Gallis K W, Araujo J T, Duff K J, Moore J G and Landry C C 1999 *Adv. Mater.* **11** 1452
 [2] Soler-Illia G J de A A, Sanchez C, Lebeau B and Patarin J 2002 *Chem. Rev.* **102** 4093
 [3] Stein A 2003 *Adv. Mater.* **15** 763 (Advances MicroMesopor Solids Stein Review)
 [4] Yang H and Zhao D 2005 *J. Chem. Mater.* **15** 1217
 [5] Lu A-H, Schüth F 2005 *C. R. Chimie* **8** 609
 [6] Hoffmann F, Cornelius M, Morell J and Fröba M 2006 *Angew. Chem. Int. Ed.* **45** 3216
 [7] Angelos S, Johansson E, Stoddart J F and Zink J I 2007 *Adv. Funct. Mater.* **17** 2261
 [8] Slowing I I, Vivero-Escoto J L, Trewyn B G and Lin V S-Y 2010 *Chem. Mater.* **22** 3790
 [9] Slowing I I, Vivero-Escoto J L, Trewyn B G and Lin V S-Y 2010 *J. Mater. Chem.* **20** 7924
[10] Li F and Stein A 2010 *Chem. Mater.* **22** 3790
[11] Linares N, Serrano E, Rico M, Balu A M, Losada E, Lugue K and García-Martínez J 2011 *Chem. Commun.* **47** 9024
[12] Nakamura T, Yamada Y and Yano K 2006 *J. Mater. Chem.* **16** 2417
[13] Li L-S, Wang Y, Young D J, Ng S-C and Tan T T Y 2010 *Electrophoresis* **31** 378
[14] Borisova D, Möhwald H and Shchukin D 2011 *ACS Nano* **5** 1939
[15] Yamada Y, Nakamura T and Yano K 2008 *Langmuir* **24** 2779
[16] Yamada Y, Nakamura T, Ishi M and Yano K 2006 *Langmuir* **22** 2444
[17] Yamada Y, Yamada H, Nakamura T and Yano K 2009 *Langmuir* **25** 13599
[18] Yamada H, Nakamura T, Yamada Y and Yano K 2009 *Adv. Mater.* **21** 4134

[19] Vivero-Escoto J L, Slowing I I, Trewyn B G, and Lin V S-Y 2010 *Small* **6** 1952
[20] Rosenholm J M, Sahlgren C and Lindén M 2010 *Nanoscale* **2** 1870
[21] Tang F, Li L and Chen D 2012 *Adv. Mater.* **24** 1504
[22] Colilla M, González B and Vallet-Regí M 2013 *Biomater. Sci.* **1** 114
[23] Milgroom A, Intrator M, Madhavan K, Mazzaro L, Shandas R, Liu B and Park D (in press) *Colloids Surf. B: Biointerfaces*
[24] Wong E W, Sheehan P E and Lieber C M 1997 *Science* **277** 1971
[25] Kuok M H, Lim H S, Ng S C, Liu N N and Wang Z K 2003 *Phys. Rev. Lett.* **90** 255502
[26] Ma R, Schliesser A, Del'Haye P, Dabirian A, Anetsberger G and Kippenberg T J 2007 *Optics lett.* **32** 2200
[27] Hashin Z and Shtrikman S 1962 *J. Mech. Phys. Solids* **10** 335
[28] Herakovich C T and Baxter S C 1999 *J. Mater. Science* **34** 1595
[29] Robertsa A P and Garboczia E J 2002 *J. Mech. Phys. Solids* **50** 33
[30] Aliev G N, Goller B and Snow P A 2011 *J. Appl. Phys.* **110** 043534
[31] Rossignol C, Chigarev N, Ducousso M, Audoin B, Forget G, Guillemot F and Durrieu M C 2008 *Appl. Phys. Lett.* **93** 123901
[32] Lee S H, Cavalieri A L, Fritz D M, Swan M C, Hegde R S, Reason M, Goldman R S and Reis D A 2005 *Phys. Rev. Lett.* **95** 246104
[33] Lin K-H, Yu C-T, Sun S-Z, Chen H-P, Pan C-C, Chyi J-I, Huang S-W, Li P-C and Sun C-K 2006 *Appl. Phys.Lett.* **89** 043106
[34] Lin K-H, Lai C-M, Pan C-C, Chyi J-I, Shi J-W, Sun S-Z, Chang C-F and Sun C-K 2007 *Nature Nanotech.* **2** 704
[35] Lomonosov A M, Ayouch A, Ruello P, Vaude G, Baklanov M R, Verdonck P, Zhao L and Gusev V E 2012 *ACS Nano* **6**, 1410
[36] Xu F, Belliarda L, Fournier D, Charron E, Duquesne J-Y, Martin S, Secouard C and Perrin B 2013 *Thin Solid Films* **548**, 366
[37] Mechri C, Ruello P and Gusev V 2012 *New Journal of Physics* **14** 023048
[38] Zerr A, Chigarev N, Brenner R, Dzivenko D A and Gusev V 2010 *Phys. Status Solidi RRL* **4** 353
[39] Trofimova E Yu, Kurdyukov D A, Kukushkina Yu A, Yagovkina M A and Golubev V G 2011 *Glass Phys. Chem.* **37** 378
[40] Trofimova E Yu, Kurdyukov D A, Yakovlev S A, Kirilenko D A, Kukushkina Yu A, Nashchekin A V, Sitnikova A A, Yagovkina M A and Golubev V G 2013 *Nanotechnology* **24** 155601
[41] Akimov A V, Tanaka Y, Pevtsov A B, Kaplan S F, Golubev V G, Tamura S, Yakovlev D R and Bayer M 2008 *Phys. Rev. Lett.* 101 033902-1-4
[42] Salasyuk A S, Scherbakov A V, Yakovlev D R, Akimov A V, Kaplyanskii A A, Kaplan S F, Grudinkin S A, Nashchekin A V, Pevtsov A B, Golubev V G, Berstermann T, Brüggem C, Bombeck M and Bayer M 2010 *Nano Lett.* **10** 1319
[43] Trofimova E Yu, Aleksenskii A E, Grudinkin S A, Korkin I V, Kurdyukov D A and Golubev V G 2011 *Colloid J.* **73** 546
[44] Grudinkin S A, Kaplan S F, Kartenko N F, Kurdyukov D A and Golubev V G 2008 *J. Phys. Chem. C* **112** 17855
[45] Kurdyukov D A, Eurov D A, Stovpiaga E Yu, Yakovlev S A, Kirilenko D A and Golubev V G 2014 *Phys. Solid State* **56** 1033
[46] Lamb H 1882 *Proc. Math. Soc. London* **13** 189
[47] Hashin Z and Shtrikman S 1962 *J. Mech. Phys. Solids* **10** 335
[48] Mechri C, Ruello P, Mounier D, Breteau J M, Povey I, Pemble M, Romanov S G and Gusev V 2007 *Journal of Physics: Conference Series* **92** 012030